\documentclass[12pt,twoside,letterpaper]{article} 
\usepackage{times,fancyhdr}
\usepackage[dvips]{graphicx}

\usepackage{amsmath}
\usepackage{amsfonts}
	
\usepackage{epstopdf}	

\usepackage{amssymb}
\usepackage{times,fancyhdr}

\begin{document}

\newcommand{\be}{\begin{equation}}
\newcommand{\ee}{\end{equation}}
\def\bq{\begin{eqnarray}}
\def\eq{\end{eqnarray}}


\title{{\bfseries  Lanczos potential of Weyl field: interpretations and applications}}
\author{\bfseries\itshape Ram Gopal Vishwakarma\thanks{E-mail address:vishwa@uaz.edu.mx}\\
 Unidad Acad$\acute{e}$mica de Matem$\acute{a}$ticas\\
 Universidad Aut$\acute{o}$noma de Zacatecas,\\
  Zacatecas, ZAC, Mexico
}

\date{}
\maketitle
\thispagestyle{empty}
\setcounter{page}{1}
\begin{abstract}
An attempt is made to uncover the physical meaning and significance of  the obscure Lanczos tensor field which is regarded as a potential of the Weyl field. Despite being a fundamental building block of any metric theory of gravity, the Lanczos tensor has not been paid proper attention as it deserves.   
By providing an elucidation on this tensor field through its derivation in some particularly chosen spacetimes, we 
try to find its adequate interpretation.

Though the Lanczos field is traditionally introduced as a gravitational analogue of  the electromagnetic 4-potential field, the performed study unearths its another feature - a relativistic analogue of the Newtonian gravitational force field. 
A new domain of applicability of the Lanczos tensor is introduced which corroborates this new feature of the tensor.
\\



\end{abstract}

\pagestyle{fancy}
\fancyhead{}
\fancyhead[EC]{Ram Gopal Vishwakarma}
\fancyhead[EL,OR]{\thepage}
\fancyhead[OC]{Lanczos potential of Weyl field: interpretations and applications}
\fancyfoot{}
\renewcommand\headrulewidth{0.5pt}

\section{Introduction}

General Relativity (GR), which constitutes the current description of gravitation in modern physics,
appears to be afflicted with at least two major difficulties. Firstly, it shows an
intrinsic difficulty in its unification with the rest of physics, which is perhaps the most profound foundational problem in physics. Secondly, its cosmological application - the $\Lambda$CDM model -
faces challenges on both theoretical and observational grounds, despite its overall success and simplicity.
It is now claimed that the current observations show evidence of physics beyond the $\Lambda$CDM model \cite{Shoes}.
This is an alarming signal to scrutinize the very foundations of the theory.
It seems that out of the four fundamental interactions, the most familiar to us is the most mysterious one.
Perhaps the present description of this interaction is incomplete just because we are not taking into account the whole system of elements responsible for the structure of spacetime.

Science thrives on crisis. Hard times call for new ideas and insights.
In this view, it would be worthwhile to draw attention towards an unnoticed ingredient of the spacetime, which is impregnated with extraordinary scientific and philosophical values and is expected to contribute to our understanding of the fundamental nature of gravitation itself.
This missing link is a rank three tensor - the Lanczos tensor $L_{\mu\nu\sigma}$ - appearing as a potential for the Weyl tensor (which represents the true gravitational degrees of freedom).
As we know, the Riemann curvature tensor plays the nodal point for the unfolding of gravity in the spacetime manifold.
However, strangely enough, the full curvature tensor  has been pushed into background. What comes forth, through the field equations of gravitation, is only the trace of the Riemann tensor (the Ricci tensor and Ricci scalar), due to its relation to the energy-momentum tensor of matter. This has led to a certain eclipse of the full curvature tensor.
It is the Lanczos tensor which fulfills this gap by providing the trace-free part of the Riemann tensor (the Weyl tensor) -  exactly those components of the tensor that are not embraced by the field equations. Thus, the understanding of the spacetime structure, and hence that of gravity, cannot be complete without the Lanczos tensor field.

Interestingly the Lanczos tensor, discovered by 
Cornelius Lanczos  in 1962 \cite{Lanczos},  happens to exist, by coincidence or providence, in any Riemannian manifold with Lorentzian signature in  four dimensions. It  generates the Weyl tensor $C_{\mu\nu\sigma\rho}$ of the manifold through the Weyl-Lanczos equation\footnote{Notations adopted: The starred symbol denotes the dual operation defined by ${*N*}_{\alpha\beta\mu\nu}=\frac{1}{4}e_{\alpha\beta\rho\sigma}e_{\mu\nu\tau\delta} N^{\rho\sigma\tau\delta}$, with $e_{\mu\nu\sigma\rho}$ representing the Levi-Civita tensor. The semicolon (comma) followed by an index denotes covariant (ordinary) derivative with respect to the corresponding variable. (Sometimes we shall also use $\nabla_\alpha$ to represent covariant derivative with respect to $x^\alpha$.) The square brackets [] enclosing indices denote skew-symmetrization, for instance, $X_{[\mu\nu]}\equiv \frac{1}{2!}(X_{\mu\nu}-X_{\nu\mu})$. Similarly,
the round brackets () enclosing indices denote symmetrization, i.e., $X_{(\mu\nu)}\equiv \frac{1}{2!}(X_{\mu\nu}+X_{\nu\mu})$. The Greek indices range over the values $0,1,2,3$ where 0 is temporal and 1, 2, 3  are spatial. For simplicity, we have considered the geometric units with $G=c=1$. 
}
\be
C_{\mu\nu\sigma\rho} = L_{[\mu\nu][\sigma;\rho]} +  L_{[\sigma\rho][\mu;\nu]} -  {*L*}_{[\mu\nu][\sigma;\rho]} -  {*L*}_{[\sigma\rho][\mu;\nu]}.   \label{eq:Weyl-Lanczos1}
\ee
The Lanczos tensor $L_{\mu\nu\sigma}$ receives  a natural and adequate interpretation in terms of the deep analogy between gravitation and electrodynamics.
In electrodynamics, a crucial ingredient of the electromagnetic field is its 4-potential $A_\mu$,  whence emanates the Faraday tensor $F_{\mu\nu}$:
\be
F_{\mu\nu} = A_{\mu;\nu} - A_{\nu;\mu}= A_{\mu,\nu} - A_{\nu,\mu},   \label{eq:F}
\ee
which measures the strength of the electromagnetic field. Let us recall that the Weyl tensor $C_{\mu\nu\sigma\rho}$ is the gravitational analogue of the electromagnetic $F_{\mu\nu}$. Hence, the relativistic potential $L_{\mu\nu\sigma}$  generating the Weyl tensor differentially, should be regarded as the gravitational analogue of the electromagnetic potential $A_\mu$ generating the field strength tensor $F_{\mu\nu}$.  This constitutes the Lanczos potential as a more fundamental geometrical object than the Weyl tensor.

The theory of this potential tensor contains many surprises and insights.
For example, it has been shown that the Minkowski spacetime also admits non-trivial values of the Lanczos potential \cite{vishwa_CQG,Takeno}, proclaiming that the Lanczos potential is a property of spacetime itself rather than  an outcome of a particular theory of gravity.
This simple but far-reaching insight of mathematical and philosophical value, revolutionizes our views on the foundational nature of spacetime itself indicating towards an important substantive nature of spacetime and its geometry.
This also seems consistent with quantum theory wherein the vacuum state of a quantum theory is not nothing. 
It would be up to the further studies to decide whether this revolutionary insight is just a mathematical curiosity or leads to a path to something deeper.
Interestingly, the Lanczos potential has already been shown to be impregnated with various signatures of quantum physics \cite{Lanczos, vishwa_CQG, N-R}.

As mentioned above, the Lanczos tensor emerges as a  fundamental geometric ingredient in any metric theory of gravity (formulated in a 4-dimensional pseudo Riemannian spacetime), irrespective of the field equations of the theory. This provides the tensor a status of  an inherent structural element in the geometric embodiment of gravity. 
Albeit its novelty and importance, this remarkable discovery is comparatively unfamiliar even
now - some sixty years after Lanczos first introduced it, and his ingenious discovery has remained more or less a mathematical curiosity. The main reason for this virtual obscurity is the absence of the physical properties of the tensor.
It has not been possible so far to ascertain what the tensor represents physically,
 although some attempts have been made recently in this direction \cite{vishwa_CQG}. 
There is also another basic question related to the Lanczos potential that needs to be answered. As the Lanczos tensor appears to be the potential of the Weyl field and the latter can be written in terms of the connections, 
then how is the  Lanczos tensor related with the connections? 

After exemplifying this tensor field by deriving its values in various spacetimes in the following sections, we
discover some possible applications thereof in some gravitational phenomena. This may help to assign an adequate physical interpretation to the tensor.

\section{Lanczos Potential Tensor and its Non-Uniqueness}
\label{sec:missing-conditions}

As Lanczos derived his tensor from a variational principle and since the latter need
not be physically consistent always, a more robust existence theorem was needed. This theorem was provided by Bampi and Caviglia, which guarantees the existence of such a potential in any
4-dimensional\footnote{While the Lanczos potential tensor does not exist in general for dimensions higher than four \cite{Edgar2000},  a potential tensor of rank 5 indeed exists in all dimensions $\geq4$. Interestingly, the trace of this potential (which amounts to its double dual) corresponds, in the case of 4-dimensions, to the usual Lanczos potential \cite{Edgar-Senovilla}.}
Riemannian manifold \cite{B-C}.
It should however be noted that  this tensor is not unique in a given spacetime.
Let us first emphasize that the tensor $L_{\mu\nu\sigma}$ satisfies the generating equation (\ref{eq:Weyl-Lanczos1})  only if it possesses the following two symmetries:
\begin{subequations}
\label{eq:SymConditions}
\begin{align}
& L_{\mu\nu\sigma} =-L_{\nu\mu\sigma},       \label{eq:syma}\\
& L_{[\mu\nu\sigma]}  =0   ~~ \Leftrightarrow   ~~ {*L}^{~~\kappa}_{\mu~~\kappa}=0.                       \label{eq:symb}
\end{align}
\end{subequations}
Let us simplify the Weyl-Lanczos equation (\ref{eq:Weyl-Lanczos1}) by evaluating the duals appearing in it, giving
\bq\nonumber
C_{\mu\nu\sigma\rho}& = & L_{\mu\nu\sigma;\rho} +  L_{\sigma\rho\mu;\nu} -  L_{\mu\nu\rho;\sigma} -  L_{\sigma\rho\nu;\mu}
+g_{\nu\sigma}L_{(\mu\rho)}+g_{\mu\rho}L_{(\nu\sigma)}\\
& - & g_{\nu\rho}L_{(\mu\sigma)}-g_{\mu\sigma}L_{(\nu\rho)}
+\frac{2}{3}L^{\lambda\kappa}_{~~~\lambda;\kappa}(g_{\mu\sigma}g_{\nu\rho}-g_{\nu\sigma}g_{\mu\rho}),\label{eq:Weyl-Lanczos}
\eq
where $L_{\mu\nu}\equiv L^{~~\kappa}_{\mu~~\nu;\kappa}-L^{~~\kappa}_{\mu~~\kappa;\nu}$.
One may check that this equation indeed admits, by virtue of the two symmetry conditions of $ L_{\mu\nu\sigma}$ given by (\ref{eq:SymConditions}), all the symmetries of the Weyl tensor: $C_{\mu\nu\sigma\rho}=C_{[\mu\nu][\sigma\rho]}$,  $C_{\mu\nu\sigma\rho}=C_{\sigma\rho\mu\nu}$,  $C_{\mu[\nu\sigma\rho]}=0$, $C^{\lambda}_{~~\mu\nu\lambda}=0$.
While the symmetry condition (\ref{eq:syma}) reduces the number of independent components of $L_{\mu\nu\sigma}$ to 24,  the condition (\ref{eq:symb}) reduces it down to 20. As the Weyl tensor has only 10 degrees of freedom,
 Lanczos considered the following additional symmetries
\begin{subequations}
\label{eq:gauge}
\begin{align}
 &  L^{~~\kappa}_{\mu~~\kappa}=0,         \label{eq:conditiona} \\
 & L^{~~~\kappa}_{\mu\nu~~;\kappa}=0,         \label{eq:conditionb}
\end{align}
\end{subequations}
as two gauge conditions in order to reduce the number of degrees of freedom present in $L_{\mu\nu\sigma}$. 
The condition (\ref{eq:conditiona}) abolishes the degeneracy in $L_{\mu\nu\sigma}$ appearing through the gauge transformation
\be
\bar{L}_{\mu\nu\sigma} = L_{\mu\nu\sigma}+ g_{\nu\sigma}X_\mu - g_{\mu\sigma}X_\nu,   \label{eq:gaugeU}
\ee
which leaves  equation (\ref{eq:Weyl-Lanczos}) invariant for an arbitrary vector field $X_\alpha$. This arbitrariness in $L_{\mu\nu\sigma}$ is fixed by the condition (\ref{eq:conditiona}) which gives $ X_\alpha=0$. 
Thus the algebraic gauge condition (\ref{eq:conditiona}), taken together with (\ref{eq:syma}) and (\ref{eq:symb}), reduces
the number of independent components of $L_{\mu\nu\sigma}$ to 16.

Let us emphasize that the differential gauge condition (\ref{eq:conditionb}), which was adopted by Lanczos just because  the divergence $ L^{~~~\kappa}_{\mu\nu~~;\kappa}$ does not participate in equation (\ref{eq:Weyl-Lanczos}),
does not further reduce the degrees of freedom of $L_{\mu\nu\sigma}$, contrary to the wide-spread misunderstanding in the literature \cite{vishwa_CQG}.
Thus ample degeneracy remains there in the values of the tensor $L_{\mu\nu\sigma}$ even after the Lanczos gauge conditions (\ref{eq:conditiona}) and (\ref{eq:conditionb}) are applied, as has been demonstrated in \cite{vishwa_CQG}.
In fact, the reason for this degeneracy in $L_{\mu\nu\sigma}$ is due to the presence of an auxiliary potential to the Weyl tensor,  discovered by Takeno \cite{Takeno}. 
He noticed that given a Lanczos potential $L_{\mu\nu\sigma}$ of a spacetime, the quantity 
\be
\bar{L}_{\mu\nu\sigma}=L_{\mu\nu\sigma}+A_{\mu\nu\sigma}\label{eq:barL}
\ee
 is again a Lanczos potential of that spacetime if  the tensor $A_{\mu\nu\sigma}$ (termed as `s-tensor' by Takeno) satisfies 
\begin{subequations}\label{eq:A}
\begin{align}
&A_{\mu\nu\sigma}=-A_{\nu\mu\sigma}, \label{eq:Aa} \\
&A_{\mu\nu\sigma}+A_{\nu\sigma\mu}+ A_{\sigma\mu\nu}=0,\label{eq:Ab} \\
&A_{[\mu\nu][\sigma;\rho]} +  A_{[\sigma\rho][\mu;\nu]} -  {*A*}_{[\mu\nu][\sigma;\rho]} -  {*A*}_{[\sigma\rho][\mu;\nu]}=0. \label{eq:Ac}
 \end{align}
\end{subequations}
It has been shown that the gauge condition (\ref{eq:conditionb}) of divergence-freeness of the Lanczos tensor, which implies that $A^{~~~\kappa}_{\mu\nu~~;\kappa}=0$,  does not fix  $A_{\mu\nu\sigma}$ uniquely and hence causes degeneracy in $L_{\mu\nu\sigma}$ \cite{vishwa_CQG}.
Thus, the cause of the degeneracy in $L_{\mu\nu\sigma}$ is the redundant degrees of freedom of the tensor, and it is not an issue with the gauge conditions. That is, the conditions  (\ref{eq:conditiona}) and (\ref{eq:conditionb}) alone, taken together with (\ref{eq:syma}) and  (\ref{eq:symb}), cannot supply a unique value of $L_{\mu\nu\sigma}$ in a given spacetime.
In order to fix this arbitrariness, one needs additional conditions/assumptions of plausibility to be imposed on $L_{\mu\nu\sigma}$, which is still missing.
In the absence of this, any arbitrary value of $L_{\mu\nu\sigma}$, just satisfying the existing conditions/symmetries, may not constitute a physically viable quantity.

Although there exists no algorithm for finding the Lanczos potential tensor in general, the tensor has indeed been found explicitly in certain special situations, for instance the spacetimes of physical significance like Schwarzschild \cite{N-V} and Kerr \cite{Bonilla}, besides many others (see \cite{review} and the references therein).
 In order to elucidate the obscure theory of Lanczos potential tensor, we derive its values in some particularly chosen spacetimes in the following. These values will also be used in the later sections while studying some possible applications of the tensor.

\subsection{Lanczos tensor for Schwarzschild Spacetime}
\label{sec:LSch}

The simplest example of the Lanczos tensor, which we take from the literature, corresponds to the Schwarzschild line element 
\be
ds^2=\left(1-\frac{2m}{r}\right) dt^2-\frac{dr^2}{(1-2m/r)}-r^2d\theta^2-r^2\sin^2\theta ~d\phi^2,\label{eq:sch}
\ee
which represents a static spherically symmetric spacetime outside an isotropic mass $m$ placed in a Ricci-flat ($R_{\mu\nu}=0$) manifold.
Novello and Velloso \cite{N-V} have shown that if a unit time-like vector field $V^\alpha\equiv dx^\alpha/ds$  tangential to the trajectory of an observer in a given spacetime is irrotational and shear-free, the Lanczos potential of the sapcetime is given by
\be
L_{\mu\nu\sigma}=V_{\mu;\lambda}V^\lambda V_\nu V_\sigma - V_{\nu;\lambda}V^\lambda V_\mu V_\sigma.\label{eq:NV}
\ee
By considering $V^\alpha=\left(\frac{1}{\sqrt{1-2m/r}},0,0,0\right)$, which comes out as irrotational and shear-free in the spacetime (\ref{eq:sch}) hence satisfying all the requirements of the Novello-Velloso formula (\ref{eq:NV}):
$V^\alpha V_\alpha=1$, shear $=$ rotation $=0$, acceleration $\neq0$. The formula (\ref{eq:NV}) then provides the Lanczos tensor with only one non-vanishing independent component
\be
L_{100}=-\frac{m}{r^2}.\label{eq:Lan1-sch}
\ee
By chance this satisfies Lanczos's gauge condition (\ref{eq:conditionb}) of divergence-freeness, though it does not satisfy the condition (\ref{eq:conditiona}) of trace-freeness. A trace-free potential can be obtained by using relation (\ref{eq:gaugeU}), which allows to cancel the trace of the tensor by choosing $X_\mu=-L_{\mu~~\lambda}^{~~\lambda}/3$ giving 
\be\label{eq:Lan2-sch}
\left.\begin{aligned}
&\bar{L}_{100}=-\frac{2m}{3r^2}\\
&\bar{L}_{122}=-\frac{m}{3(1-2m/r)}\\
&\bar{L}_{133}=-\frac{m\sin^2\theta}{3(1-2m/r)}
 \end{aligned}
 \right\},
\ee
which satisfy both Lanczos gauge conditions given in (\ref{eq:gauge}) fortunately. 
We find another solution of equation (\ref{eq:Weyl-Lanczos}), not realized so far in the literature, for the Schwarzschild line element (\ref{eq:sch}):
\be\label{eq:Lan3-sch}
\left.\begin{aligned}
&\tilde{L}_{100}=-\frac{1}{2}\frac{m}{r^2}\\
&\tilde{L}_{133}=\frac{1}{2}r\sin^2\theta\\
&\tilde{L}_{233}=\frac{1}{4}r^2\sin2\theta
 \end{aligned}
 \right\},
\ee
satisfying the gauge condition (\ref{eq:conditionb}) of the divergence freeness but not satisfying the gauge condition  (\ref{eq:conditiona}) of trace-freeness. The trace-free form can similarly be obtained by using (\ref{eq:gaugeU}), though at the cost of letting more non-vanishing components enter it.

\subsection{Lanczos Tensor for the Schwarzschild-de Sitter Spacetime}

The Lanczos potential of the Schwarzschild spacetime can easily be generalized by adding the cosmological constant
$\Lambda$ in (\ref{eq:sch}) thereby resulting in the Schwarzschild-de Sitter solution of $G_{\mu\nu}+\Lambda g_{\mu\nu}=0$:
\be
ds^2=\left(1-\frac{2m}{r}-\frac{\Lambda r^2}{3}\right) dt^2-\frac{dr^2}{(1-2m/r -\Lambda r^2/3)}-r^2d\theta^2-r^2\sin^2\theta ~d\phi^2.\label{eq:sch-deSitter}
\ee
The choice $V^\alpha=\left( \left(1-\frac{2m}{r}-\frac{\Lambda r^2}{3}\right)^{-1/2},0,0,0\right)$ satisfies all the requirements of the Novello-Velloso formula (\ref{eq:NV}) for the line element (\ref{eq:sch-deSitter}). The formula then gives
only one non-vanishing independent component
\be
L_{100}=\frac{\Lambda r}{3}-\frac{m}{r^2}\label{eq:Lan1-sch-deSitter}
\ee
satisfying the gauge condition (\ref{eq:conditionb}) but not (\ref{eq:conditiona}).
A trace-free potential is obtained by the use of (\ref{eq:gaugeU}) giving
\be\label{eq:Lan2-sch-deSitter}
\left.\begin{aligned}
&\bar{L}_{100}=\frac{2}{3}\left( \frac{\Lambda r}{3}-\frac{m}{r^2} \right)\\
&\bar{L}_{122}= \frac{\frac{\Lambda r^3}{3}-m}{3(1-2m/r-\Lambda r^2/3)}\\
&\bar{L}_{133}=\frac{\left(\frac{\Lambda r^3}{3}-m\right)\sin^2\theta}{3(1-2m/r-\Lambda r^2/3)}
 \end{aligned}
 \right\},
\ee
which satisfies both gauge conditions (\ref{eq:gauge}).

\subsection{Lanczos Tensor for the Schwarzschild Interior Spacetime}
\label{sec:Lsch-Int}

It may appear surprising that the Lanczos tensor, that appears as the potential of the Weyl tensor, may be non-vanishing even when the Weyl tensor vanishes. In order to exemplify this, let us first consider the Schwarzschild interior solution
\be
ds^2=\frac{1}{4}\left(3\sqrt{1-\frac{r_0^2}{R^2}}-\sqrt{1-\frac{r^2}{R^2}}\right)^2 dt^2-\frac{dr^2}{1-r^2/R^2}-r^2 d\theta^2-r^2\sin^2\theta ~d\phi^2,\label{eq:sch-Int}
\ee
which, being a static and isotropic solution of Einstein equation, represents the spacetime inside a non-rotating, incompressible perfect fluid sphere of constant density $\rho_0$ and radius $r_0$. Here $R\equiv \sqrt{3/8\pi \rho_0}$. As the solution is conformally flat, its Weyl tensor vanishes identically.
It would not be difficult to check that a $V^\alpha=\left(\left(\frac{3}{2}\sqrt{1-\frac{r_0^2}{R^2}}-\frac{1}{2}\sqrt{1-\frac{r^2}{R^2}}\right)^{-1},0,0,0\right)$ in the spacetime (\ref{eq:sch-Int}), satisfies all the requirements of the Novello-Velloso formula (\ref{eq:NV}):
$V^\alpha V_\alpha=1$, shear $=$ rotation $=0$, acceleration $\neq0$. The formula then gives
\be
L_{100}= \frac{5 r\left[ 1-\frac{r^2+9r_0^2}{10R^2} -\frac{3}{5}\sqrt{\left(1-\frac{r^2}{R^2}\right)\left(1-\frac{r_0^2}{R^2}\right)}\right]}{2 R^2\left[ \left(1-\frac{r^2}{R^2}\right)-3\sqrt{\left(1-\frac{r^2}{R^2}\right)\left(1-\frac{r_0^2}{R^2}\right)}\right]}\label{eq:Lan1-sch-Int}
\ee
as the only non-vanishing independent component of $L_{\mu\nu\sigma}$, which is divergence-free but not trace-free.
A trace-free form can similarly be obtained by the use of (\ref{eq:gaugeU}), giving
\be\label{eq:Lan2-sch-Int}
\left.\begin{aligned}
&\bar{L}_{100}= \frac{5 r\left[ 1-\frac{r^2+9r_0^2}{10R^2} -\frac{3}{5}\sqrt{\left(1-\frac{r^2}{R^2}\right)\left(1-\frac{r_0^2}{R^2}\right)}\right]}{3 R^2\left[ \left(1-\frac{r^2}{R^2}\right)-3\sqrt{\left(1-\frac{r^2}{R^2}\right)\left(1-\frac{r_0^2}{R^2}\right)}\right]}\\
&\bar{L}_{122}= \frac{10 r^3\left[ 1-\frac{r^2+9r_0^2}{10R^2} -\frac{3}{5}\sqrt{\left(1-\frac{r^2}{R^2}\right)\left(1-\frac{r_0^2}{R^2}\right)}\right]}{3 R^2\left[ \left(1-\frac{r^2}{R^2}\right)^{2/3}-3\left(1-\frac{r^2}{R^2}\right)^{1/6}\left(1-\frac{r_0^2}{R^2}\right)^{1/2}\right]^3}\\
&\bar{L}_{133}=\frac{10 r^3\left[ 1-\frac{r^2+9r_0^2}{10R^2} -\frac{3}{5}\sqrt{\left(1-\frac{r^2}{R^2}\right)\left(1-\frac{r_0^2}{R^2}\right)}\right]}{3 R^2\left[ \left(1-\frac{r^2}{R^2}\right)^{2/3}-3\left(1-\frac{r^2}{R^2}\right)^{1/6}\left(1-\frac{r_0^2}{R^2}\right)^{1/2}\right]^3} \sin^2\theta
 \end{aligned}
 \right\},
\ee
which satisfies both gauge conditions documented in (\ref{eq:gauge}).

Let us note that if $L_{\mu\nu\sigma}$ is a Lanczos potential of a conformally flat (or flat) spacetime, then for an arbitrary constant $\kappa$, the tensor $\kappa L_{\mu\nu\sigma}$  is also its Lanczos potential by virtue of the vanishing Weyl tensor  and the linearity of equation (\ref{eq:Weyl-Lanczos})  in $L_{\mu\nu\sigma}$.

\subsection{Lanczos Tensor for the Robertson-Walker Spacetimes}
\label{sec:LRW}

As another example of a spacetime with vanishing Weyl tensor, let us consider the Robertson-Walker (R-W) line element (which too is conformally flat) in the form
\be
ds^2=dt^2-S^2(t)\left[\frac{dr^2}{1-\alpha r^2}+r^2(d\theta^2+\sin^2\theta ~d\phi^2)\right],\label{eq:RW}
\ee
where $\alpha$ has been kept as an arbitrary constant (not normalized, as in the usual standard form) to be used later  in section \ref{sec:CollapsingStar}. 
The Lanczos tensor for some particular cases of this spacetime has been discovered in \cite{vishwa_CQG}. Here we discover its value in a more general case. For this purpose, we follow a heuristic approach and take guidelines from the Schwarzschild case (\ref{eq:sch}). Since both the line elements (\ref{eq:sch}) and (\ref{eq:RW}) are spherically symmetric, we expect that it should be possible to have $L_{\mu\nu\sigma}$ for the R-W line element (\ref{eq:RW}) with a single (independent) non-vanishing component, like the one given by (\ref{eq:Lan1-sch}) for the Schwarzschild line element (\ref{eq:sch}). But apart from the radial coordinate $r$, now  in the case of the line element (\ref{eq:RW}) we should also expect contributions from $S$ and possibly from $(1-\alpha r^2)$ also. Hence, we expect 
\be
L_{100}=\kappa r^\ell S^n (1-\alpha r^2)^p,  \label{eq:Lan1-RW}
\ee
to form the only independent non-vanishing component of the Lanczos potential for the spacetime (\ref{eq:RW}) for suitable values of the parameters $\ell,n$ and $p$. These parameters are to be determined by the condition that (\ref{eq:Lan1-RW}) should satisfy the Weyl-Lanczos generating equation (\ref{eq:Weyl-Lanczos}) for the line element (\ref{eq:RW}) with $C_{\mu\nu\sigma\rho}=0$. Let us recall that the spacetime (\ref{eq:RW}) is conformally flat and hence its Weyl tensor vanishes identically. The multiplicative constant $\kappa$ has been introduced in (\ref{eq:Lan1-RW}) for the same reason, as explained earlier, that if $L_{\mu\nu\sigma}$ is a solution of (\ref{eq:Weyl-Lanczos}) for the line element (\ref{eq:RW}), then so is $\kappa L_{\mu\nu\sigma}$.

If we denote the right hand side of equation (\ref{eq:Weyl-Lanczos}) by $W_{\mu\nu\sigma\rho}$, then its non-vanishing components generated by the definition (\ref{eq:Lan1-RW}) for the spacetime (\ref{eq:RW}) yield
\be\label{eq:W}
\left.\begin{aligned}
& W_{1010}=\frac{2 \kappa}{3} r^{\ell-1}(1-\alpha r^2)^{p-1}S^{n}[1-\ell+\alpha r^2(\ell+2p)],\\
& W_{1212}=\frac{\kappa}{3} r^{\ell+1}(1-\alpha r^2)^{p-1}S^{n+2}[1-\ell+\alpha r^2(\ell+2p)],\\
& W_{1313}=\frac{\kappa}{3} r^{\ell+1}(1-\alpha r^2)^{p-1}S^{n+2}[1-\ell+\alpha r^2(\ell+2p)]\sin^2\theta,\\
& W_{2020}=-\frac{\kappa}{3} r^{\ell+1}(1-\alpha r^2)^p ~S^n[1-\ell+\alpha r^2(\ell+2p)],\\
& W_{3030}=-\frac{\kappa}{3} r^{\ell+1}(1-\alpha r^2)^p ~S^n[1-\ell+\alpha r^2(\ell+2p)]\sin^2\theta,\\
& W_{2323}=-\frac{2\kappa}{3} r^{\ell+3}(1-\alpha r^2)^{p}~S^{n+2}[1-\ell+\alpha r^2(\ell+2p)]\sin^2\theta.
 \end{aligned}
 \right\}
\ee
Obviously, the vanishing of all these components simultaneously (for all $r$, $S$ and $\kappa\neq0$), requires 
\[
1-\ell+\alpha r^2(\ell+2p)=0,
\]
which gives a unique solution 
\[
\ell=1,~~~~p=-\frac{1}{2},
\]
to be compatible with an arbitrary value of $\alpha$. Thus 
\be
L_{100}=\kappa \frac{r S^n}{\sqrt{ 1-\alpha r^2}},  \label{eq:Lan2-RW}
\ee
constitutes the Lanczos potential for the spacetime (\ref{eq:RW}) with $\kappa$, $n$ and $\alpha$  as arbitrary constants.
The trace $ L^{~~\lambda}_{\mu~~\lambda}$  and the divergence $L^{~~~\lambda}_{\mu\nu~~;\lambda}$ of the tensor can be calculated, giving the non-vanishing components as
\be
L^{~~\lambda}_{1~~\lambda}=\kappa \frac{r S^n}{\sqrt{ 1-\alpha r^2}},
\ee 
\be
 L^{~~~\lambda}_{10~~;\lambda}=-L^{~~~\lambda}_{01~~;\lambda}=\kappa (n+2)\frac{r S^{n-1}\dot{S}}{\sqrt{ 1-\alpha r^2}}.\label{eq:divergence}
\ee
Clearly the solution (\ref{eq:Lan2-RW}) does not satisfy the condition (\ref{eq:conditiona}) [nor does it satisfy the condition (\ref{eq:conditionb})].
As we have mentioned earlier, a trace-free potential can be obtained by using relation (\ref{eq:gaugeU}), which allows to cancel the trace of the tensor by choosing $X_\mu=-L_{\mu~~\lambda}^{~~\lambda}/3$ giving now the following three independent non-vanishing components:
\be\label{eq:Lan3-RW}
\left.\begin{aligned}
& L_{100}=\frac{2\kappa}{3}\frac{r S^n}{\sqrt{1-\alpha r^2}},\\
& L_{122}=\frac{\kappa}{3}\frac{r^3 S^{n+2}}{\sqrt{1-\alpha r^2}},\\
& L_{133}=\frac{\kappa}{3}\frac{r^3 S^{n+2}}{\sqrt{1-\alpha r^2}}\sin^2\theta.
 \end{aligned}
 \right\}
\ee
 This can be generalized into yet another solution
\be\label{eq:Lan4-RW}
\left.\begin{aligned}
& L_{100}=\kappa_1\frac{r S^n}{\sqrt{1-\alpha r^2}},\\
& L_{122}=\kappa_2\frac{r^3 S^{n+2}}{\sqrt{1-\alpha r^2}},\\
& L_{133}=\kappa_2\frac{r^3 S^{n+2}}{\sqrt{1-\alpha r^2}}\sin^2\theta,
 \end{aligned}
 \right\}
\ee
though at the cost of loosing the gauge symmetry (\ref{eq:conditiona}). Here $\kappa_1, \kappa_2$ are arbitrary constants.

\bigskip

Our elucidation on the Lanczos tensor is now complete to some extent, and it is ready to be involved in applications. 
Although much attention has been devoted to the Weyl tensor in the  research on a gravitational theory, the same did not occur with the Lanczos potential.
Since the Lanczos tensor appears as a fundamental property of spacetime itself irrespective of any particular theory of gravitation, it is expected to be imbued with more interesting physical properties than the Weyl tensor. With the values of this tensor derived in these sections, we explore what physical information this potential tensor may convey 
 in a physical situation.

\section{Conformal Properties of Lanczos Tensor}

Conformal symmetry  plays an important role in field theories including gravity and leads to interesting insights about the physical phenomena related to the dynamics of spacetime.  Under this symmetry for instance, all the features of gravitational wave propagation and the global causal structure of spacetime  remain conserved, as the conformal transformations leave the null geodesics and hence the light cones invariant. 

A conformal transformation is a rescaling of the space and time intervals performed without changing the coordinates used to describe events in a  spacetime manifold. This rescaling is brought about through the metric transformation
\be
g_{\mu \nu} \rightarrow \tilde{g}_{\mu \nu}=\Omega^2 g_{\mu \nu},\label{eq:conformal}
\ee
where the conformal factor $\Omega(x^\alpha)$ is a dimensionless, positive and smooth but otherwise arbitrary function of the spacetime coordinates. This gives rise to a new spacetime and its geometry characterized by the metric $\tilde{g}_{\mu \nu}$, wherein the inverse metric and the Christoffel symbols lead to
\begin{subequations}
\begin{align}
&\tilde{g}^{\mu \nu}=\Omega^{-2} g^{\mu \nu},\label{eq:InvConformal}\\
&\tilde{\Gamma}^\alpha_{\beta \gamma}=\Gamma^\alpha_{\beta \gamma}+\Omega^{-1} \left(\delta^\alpha_\beta ~\Omega_{,\gamma}+\delta^\alpha_\gamma ~\Omega_{,\beta}-g_{\beta \gamma}~g^{\alpha\lambda}~\Omega_{,\lambda}\right).\label{eq:Christof}
\end{align}
\end{subequations}
The tilde over a quantity is going to denote its value in the conformally rescaled spacetime with metric $\tilde{g}_{\mu \nu}$. As the Weyl tensor is fully invariant under the transformation (\ref{eq:conformal}), it is natural to ask how this conformal invariance is manifested in the Lanczos tensor which plays the role of the potential for the Weyl tensor.
But, unlike the Weyl tensor, it has not so far been possible to obtain an expression for the Lanczos tensor in terms of the metric tensor so that its conformal invariance can be studied directly. Nevertheless one can still analyze the conformal invariance of the Weyl-Lanczos equation (\ref{eq:Weyl-Lanczos}) by considering a {\it conformal weight} for $L_{\mu\nu\sigma}$:
\be
\tilde{L}_{\mu\nu\sigma}=\Omega^s L_{\mu\nu\sigma},\label{eq:conformalL}
\ee
where $s$ is a real number  (see the Appendix D of Reference \cite{Wald}).
If we denote the  right hand side of equation (\ref{eq:Weyl-Lanczos}) with $W_{\mu\nu\sigma\rho}$ as we have done earlier, then the application of the rescaling (\ref{eq:conformalL}) must render  $W^\mu_{~~\nu\sigma\rho}$ unchanged under the transformation (\ref{eq:conformal}), i.e. $\tilde{W}^\mu_{~~\nu\sigma\rho}=W^\mu_{~~\nu\sigma\rho}$, since so is the left hand side of equation (\ref{eq:Weyl-Lanczos}): $\tilde{C}^\mu_{~~\nu\sigma\rho}=C^\mu_{~~\nu\sigma\rho}$. 

In order to look into this matter directly, let us compute the right hand side of equation (\ref{eq:Weyl-Lanczos}) in the conformally transformed spacetime.
Use of the conformal rescalings (\ref{eq:InvConformal}, \ref{eq:Christof}, \ref{eq:conformalL}) and a long  but straightforward calculation produce
\be\label{eq:W-LConformal}
\left.\begin{aligned}
\tilde{W}^\mu_{~~\nu\sigma\rho}=\Omega&^{(s-2)}  W^\mu_{~~\nu\sigma\rho}\\
&+\frac{(s-3)}{2}\Omega^{(s-3)}g^{\mu\alpha}\big[2\big\{L_{\alpha\nu\sigma}~\Omega_{,\rho}+L_{\nu\alpha\rho}~\Omega_{,\sigma}+ L_{\sigma\rho\alpha}~\Omega_{,\nu}+L_{\rho\sigma\nu}~\Omega_{,\alpha} \big\}\\
&+\big\{g_{\alpha\rho}(L^{~\kappa}_{\nu ~ \sigma}+L^{~\kappa}_{\sigma ~ \nu})
+ g_{\nu\sigma}(L^{~\kappa}_{\alpha ~ \rho}+L^{~\kappa}_{\rho ~ \alpha})- g_{\alpha\sigma}(L^{~\kappa}_{\nu ~ \rho}+L^{~\kappa}_{\rho ~ \nu})- g_{\nu\rho}(L^{~\kappa}_{\alpha ~ \sigma}+L^{~\kappa}_{\sigma ~ \alpha})  \big\}\Omega_{,\kappa}\\
&+\big\{\big(g_{\nu\rho}L^{~\kappa}_{\sigma ~ \kappa}-g_{\nu\sigma}L^{~\kappa}_{\rho ~ \kappa}\big)\Omega_{,\alpha}+
\big(g_{\alpha\sigma}L^{~\kappa}_{\rho ~ \kappa}-g_{\alpha\rho}L^{~\kappa}_{\sigma ~ \kappa}\big)\Omega_{,\nu}+
\big(g_{\rho\nu}L^{~\kappa}_{\alpha ~ \kappa}-g_{\rho\alpha}L^{~\kappa}_{\nu ~ \kappa}\big)\Omega_{,\sigma}\\
&~~~~~~~~~~~~~~~~~~~~~~~~+\big(g_{\sigma\alpha}L^{~\kappa}_{\nu ~ \kappa}-g_{\sigma\nu}L^{~\kappa}_{\alpha ~ \kappa}\big)\Omega_{,\rho}\big\}\\
&- \frac{4}{3}L^{\lambda\kappa}_{~ ~ ~\kappa}~\Omega_{,\lambda}\big(g_{\alpha\sigma}~g_{\nu\rho}-g_{\nu\sigma}~g_{\alpha\rho}\big)\big].
 \end{aligned}
\right.
\ee
Thus in general the potential tensor $L_{\mu\nu\sigma}$, defined through  (\ref{eq:Weyl-Lanczos}), fails to produce $\tilde{W}^\mu_{~~\nu\sigma\rho}=W^\mu_{~~\nu\sigma\rho}$ if the tensor is limited to admit the symmetries (\ref{eq:SymConditions}) only. This is due to the presence of the second term on the right hand side of  equation (\ref{eq:W-LConformal}). Although this undesired term vanishes for $s=3$, this choice does not fulfill the requirement $\tilde{W}^\mu_{~~\nu\sigma\rho}=W^\mu_{~~\nu\sigma\rho}$. The only way to achieve this goal is to have $s=2$ together with imposing a condition on $L_{\mu\nu\sigma}$ given by
\be\label{eq:ConfCondition}
\left.\begin{aligned}
2\big\{&L_{\alpha\nu\sigma}~\Omega_{,\rho}+L_{\nu\alpha\rho}~\Omega_{,\sigma}+ L_{\sigma\rho\alpha}~\Omega_{,\nu}+L_{\rho\sigma\nu}~\Omega_{,\alpha} \big\}
+\big\{g_{\alpha\rho}(L^{~\kappa}_{\nu ~ \sigma}+L^{~\kappa}_{\sigma ~ \nu})
+ g_{\nu\sigma}(L^{~\kappa}_{\alpha ~ \rho}+L^{~\kappa}_{\rho ~ \alpha})\\
&- g_{\alpha\sigma}(L^{~\kappa}_{\nu ~ \rho}+L^{~\kappa}_{\rho ~ \nu})- g_{\nu\rho}(L^{~\kappa}_{\alpha ~ \sigma}+L^{~\kappa}_{\sigma ~ \alpha})  \big\}\Omega_{,\kappa}
+\big\{\big(g_{\nu\rho}L^{~\kappa}_{\sigma ~ \kappa}-g_{\nu\sigma}L^{~\kappa}_{\rho ~ \kappa}\big)\Omega_{,\alpha}\\
&+\big(g_{\alpha\sigma}L^{~\kappa}_{\rho ~ \kappa}-g_{\alpha\rho}L^{~\kappa}_{\sigma ~ \kappa}\big)\Omega_{,\nu}+
\big(g_{\rho\nu}L^{~\kappa}_{\alpha ~ \kappa}-g_{\rho\alpha}L^{~\kappa}_{\nu ~ \kappa}\big)\Omega_{,\sigma}
+\big(g_{\sigma\alpha}L^{~\kappa}_{\nu ~ \kappa}-g_{\sigma\nu}L^{~\kappa}_{\alpha ~ \kappa}\big)\Omega_{,\rho}\big\}\\
&- \frac{4}{3}L^{\lambda\kappa}_{~ ~ ~\kappa}~\Omega_{,\lambda}\big(g_{\alpha\sigma}~g_{\nu\rho}-g_{\nu\sigma}~g_{\alpha\rho}\big)=0.
 \end{aligned}
\right.
\ee
It is not clear at this point what kind of symmetry this equation imposes on $L_{\mu\nu\sigma}$ in addition to the two obligatory symmetries (\ref{eq:syma}) and (\ref{eq:symb}). In the following we discuss a simple possibility which is suggested by breaking the long equation (\ref{eq:ConfCondition}) in four parts according to the symmetries of the terms:
\begin{subequations}
\label{eq:ConCons}
\begin{align}
&L_{\alpha\nu\sigma}~\Omega_{,\rho}+L_{\nu\alpha\rho}~\Omega_{,\sigma}+ L_{\sigma\rho\alpha}~\Omega_{,\nu}+L_{\rho\sigma\nu}~\Omega_{,\alpha} =0,  \label{eq:ConCona}\\
&\big\{g_{\alpha\rho}(L^{~\kappa}_{\nu ~ \sigma}+L^{~\kappa}_{\sigma ~ \nu})
+ g_{\nu\sigma}(L^{~\kappa}_{\alpha ~ \rho}+L^{~\kappa}_{\rho ~ \alpha})- g_{\alpha\sigma}(L^{~\kappa}_{\nu ~ \rho}+L^{~\kappa}_{\rho ~ \nu})- g_{\nu\rho}(L^{~\kappa}_{\alpha ~ \sigma}+L^{~\kappa}_{\sigma ~ \alpha})  \big\}\Omega_{,\kappa}=0,\label{eq:ConConb}\\\nonumber
&\big(g_{\nu\rho}L^{~\kappa}_{\sigma ~ \kappa}-g_{\nu\sigma}L^{~\kappa}_{\rho ~ \kappa}\big)\Omega_{,\alpha}
+\big(g_{\alpha\sigma}L^{~\kappa}_{\rho ~ \kappa}-g_{\alpha\rho}L^{~\kappa}_{\sigma ~ \kappa}\big)\Omega_{,\nu}+
\big(g_{\rho\nu}L^{~\kappa}_{\alpha ~ \kappa}-g_{\rho\alpha}L^{~\kappa}_{\nu ~ \kappa}\big)\Omega_{,\sigma}\\
&+\big(g_{\sigma\alpha}L^{~\kappa}_{\nu ~ \kappa}-g_{\sigma\nu}L^{~\kappa}_{\alpha ~ \kappa}\big)\Omega_{,\rho}=0,\label{eq:ConConc}\\
& L^{\lambda\kappa}_{~ ~ ~\kappa}~\Omega_{,\lambda}=0.\label{eq:ConCond}
\end{align}
\end{subequations}
The last equation, taken together with the arbitrariness in $\Omega$, implies that the Lanczos potential tensor $L_{\mu\nu\sigma}$ should be trace-free:
\be
L^{~\kappa}_{\mu ~ \kappa}=0. \label{eq:ConCone}
\ee
This not only satisfies identically the condition (\ref{eq:ConConc}) but also appears consistent with the general tenet that the conformally invariant tensors must be trace-free, thus contributing further to the aesthetic appeal of (\ref{eq:ConCone}). Let us note that the scaling law (\ref{eq:conformalL}), for $s=2$, implies that the tensor $L_{\mu\nu\sigma}$ (with one index raised) is conformally invariant: $\tilde{L}^\mu_{~~\nu\sigma}=L^\mu_{~~\nu\sigma}$, akin to $\tilde{C}^\mu_{~~\nu\sigma\rho}=C^\mu_{~~\nu\sigma\rho}$.  (However, this is not so with the tensor with all covariant indices:  $\tilde{L}_{\mu\nu\sigma}=\Omega^2 L_{\mu\nu\sigma}$, akin to $\tilde{C}_{\mu\nu\sigma\rho}=\Omega^2 C_{\mu\nu\sigma\rho}$.)

Thus under this simple case, the constraints (\ref{eq:ConCona}), (\ref{eq:ConConb}) and (\ref{eq:ConCone}) are imposed on the Lanczos tensor by the conformal invariance of the Weyl tensor (which must be met) brought about through the Weyl-Lanczos equation. This may supply the missing plausibility conditions to be imposed on the Lanczos tensor as has been mentioned in section \ref{sec:missing-conditions}. Further study is required to decipher the meaning of the more general case (\ref{eq:ConfCondition}).

\section{Possible Interpretations of Lanczos Tensor}

We have derived the Lanczos potential tensor in various cases of spacetime in the above examples.
However, unlike the Weyl tensor, it has not so far been possible to obtain a general expression for this tensor in terms of the metric tensor. Though this expression does exist in the case of a weak gravitational field, wherein
the spacetime metric differs minutely from the Minkowskian metric $\eta_{\mu\nu}$:
\be
g_{\mu\nu}= \eta_{\mu\nu}+ h_{\mu\nu},~~~~~~{\rm where} ~~~~|h_{\mu\nu}|<<1.\label{eq:weak}
\ee
In this case, the Lanczos tensor can be written as \cite{Lanczos}
\be
L_{\mu\nu\sigma}=\frac{1}{4}\left(h_{\mu\sigma,\nu} - h_{\nu\sigma,\mu} +\frac{1}{6} h_{,\mu}\eta_{\nu\sigma} - \frac{1}{6} h_{,\nu}\eta_{\mu\sigma}\right),  \label{eq:weakL}
\ee 
 to linear order in the metric perturbations $h_{\mu\nu}$. Here $h\equiv h_{\mu\nu}\eta^{\mu\nu}$.

\subsection{Minkowskian Lanczos Tensor as a Ground State `Potential'}

Let us note that the covariant and ordinary differentiations coincide in the first order of approximation in the case of a weak gravitational field  and hence the differential and algebraic tensor operations are performed by the use of the  Minkowskian metric $\eta_{\mu\nu}$, as is the case with $L_{\mu\nu\sigma}$ defined by (\ref{eq:weakL}). This is exactly the case with the Lanczos potential for the  Minkowski spacetime, which has been elaborated on in \cite{vishwa_CQG, Takeno}\footnote{The Lanczos tensor of any conformally flat spacetime including the flat spacetime itself can be defined by equations (\ref{eq:Aa}-\ref{eq:Ac}) (with  the understanding of $A_{\mu\nu\sigma}$ denoting the Lanczos tensor of the corresponding spacetime).}. This implies that  if $L_{\mu\nu\sigma}$ is the Lanczos potential of a given spacetime in a weak field, then so is $L_{\mu\nu\sigma}+\overset{\rm{\scriptscriptstyle GS}}{L}_{\mu\nu\sigma}$, where $\overset{\rm{\scriptscriptstyle GS}}{L}_{\mu\nu\sigma}$ is the Lanczos potential of
the Minkowskian spacetime. Thus $\overset{\rm{\scriptscriptstyle GS}}{L}_{\mu\nu\sigma}$
 appears as a `ground state' potential field which makes an essential contribution to the Lanczos potential fields of all the curved  spacetimes  in a weak gravitational field.
This provides the Minkowskian Lanczos potential  a natural and adequate interpretation portraying it as the `weight'  or  the `metrical elasticity' of spacetime, which being an integral part of all the spacetime geometries, opposes the curving of spacetime.

Assigning a `ground state'  potential field to the Minkowskian spacetime in the absence of any curvature, may appear puzzling and surprising at the first glance. However, there does exist a parallel to this situation in the quantum realm wherein one can have a vanishing electromagnetic field in a region but a non-vanishing potential, explaining  the conspicuous physical effects which do exist in that region (Aharonov-Bohm effect) \cite{vishwa_CQG}. 

A word of caution is in order here. It should be noted that the Lanczos `potential' does not have the dimensions of the potential energy, in the same way as the Weyl tensor does not have the dimensions of force. Thus the denomination of the Lanczos tensor as the `Lanczos potential', refers to just the electromagnetic analogy.
Let us recall that the Lanczos tensor is regarded as the gravitational analogue of the electromagnetic potential, since it generates the Weyl tensor differentially, in the same way as the electromagnetic potential generates the Faraday tensor. Hence, the  Lanczos tensor is generally termed as the potential to the Weyl tensor.

\subsection{Lanczos Tensor as a Relativistic Analogue of the Newtonian Force}
\label{sec:ForceAnalogy}

Since the Lanczos tensor generates the Weyl tensor and since the latter is linked with the (free) gravitational field, one may naturally expect that the Lanczos tensor too has something to do with the gravitational field. In this view,  the quantity $L_{\mu\nu\sigma}$ given by expression (\ref{eq:weakL}) in terms of the derivatives of the metric tensor, gives a clue that  the Lanczos tensor may represent a relativistic analogue of the Newtonian force. However if this is so, the tensor should reduce to the Newtonian gravitational force in a physical situation 
 in the weak field and low velocity limit.
As the Newtonian theory of gravitation provides excellent approximations under a wide range of
astrophysical cases, the first crucial test of any theory of gravitation is that it reduces to the Newtonian
gravitation.

In order to check this, let us consider a static spherical mass $m$ placed at the origin of a centrally symmetric coordinate system $r,\theta,\phi$.
In the Newtonian theory of gravity, the gravitational field produced by the mass at a point $r$ is represented in terms of the gravitational potential $\Phi(r)=-m/r$ at that point.
 In a relativistic theory of gravitation, for example GR, the gravitational field of the mass is well-described by the Schwarzschild line element (\ref{eq:sch}). In a weak field,
the line element (\ref{eq:sch}) reduces to
\be
ds^2=\eta_{\mu\nu}dx^\mu dx^\nu -\frac{2m}{r}(dt^2+dr^2),\label{eq:schW}
\ee
at large $r$ so that $2m/r<<1$. Its comparison with the condition (\ref{eq:weak}) gives the only non-vanishing components of $h_{\mu\nu}$ as $h_{00}=h_{11}=-2m/r$.
 For this, the definition (\ref{eq:weakL}) provides 
\be
L_{100}=-\frac{m}{2r^2},  \label{eq:schwWL}
\ee
as the only non-vanishing independent component of $L_{\mu\nu\sigma}$, which amounts to
\be
L_{100}=-\frac{1}{2}\frac{\partial \Phi}{\partial r},  \label{eq:schwWL1}
\ee
by virtue of the well-known relation $g_{00}=(1+2\Phi)$ valid in the weak field and low velocity limit. This can be written in terms of $F_{\rm N}$, the corresponding Newtonian force experienced by a unit test mass placed at a distance $r$ from the source mass:
\be
L_{100}=\frac{1}{2}F_{\rm N}.  \label{eq:schwWL2}
\ee
 Thus the Lanczos tensor indeed represents the relativistic analogue of the Newtonian gravitational force in a curved spacetime in the weak field and low velocity limit (the factor $\frac{1}{2}$ can be considered as a relativistic effect). It may be mentioned that conventionally it is the Christoffel symbol which is considered as the analogue of the Newtonian force. However, as the Christoffel symbols are not tensors, the Lanczos tensor suits better to a covariant theory in representing the analogue of the Newtonian force when the analogue of the Newtonian potential is ascribed to a tensor (the metric tensor). 

This new understanding of the Lanczos tensor also makes a prediction: the tensor should match at the boundary of the two spacetime regions on the surface of a gravitating body.
In order to exemplify this, let us apply the force analogy on a perfect fluid ball of constant density wherein the spacetime inside the ball is given by the Schwarzschild interior line element (\ref{eq:sch-Int}), and that outside the ball by the Schwarzschild (exterior) line element (\ref{eq:sch}). According to the force analogy, the Lanczos tensor would approximate the force experienced by a unit test mass placed on the surface of the ball. 
But the Lanczos tensor can be calculated from either of the two line elements, which of course match at the boundary. It is then expected that the Lanczos tensor too should match smoothly at the boundary, if the force analogy is correct. This is indeed the case, as we shall witness in section \ref{sec:StaticStar}. This scenario is also corroborated by the case of the collapsing ball discussed in section \ref{sec:CollapsingStar}.

This analogy however also poses a question: How is the gravitational information encoded in the Lanczos tensor?
It is clear that in the above-described example, the Newtonian force on the test mass is estimated in terms of the mass of the gravitating ball, which enters the scene through the Einstein field equation. In other words, the information on this force (and hence that on matter and gravity) gets encoded in Lanczos tensor through Einstein's tensor, i.e. through Ricci's tensor and scalar.
Let us also recall that the Riemann tensor is decomposed in terms of its two mutually independent parts - its trace-free part (the Weyl tensor) and its trace (the Ricci tensor and Ricci scalar) as
\be
R_{\mu\nu\sigma\rho}=C_{\mu\nu\sigma\rho} - g_{\mu[\rho}R_{\sigma]\nu} - g_{\nu[\sigma}R_{\rho]\mu} - \frac{1}{3}Rg_{\mu[\sigma}~g_{\rho]\nu}.\label{eq:R-W-R}
\ee
As the energy-momentum tensor of matter is correlated with the trace of Riemann tensor (through the field equation), equation (\ref{eq:R-W-R}) indicates that the Weyl tensor is algebraically independent of the matter tensor.
Hence so is the Lanczos tensor (which is the potential of Weyl tensor), since the Ricci's tensor and scalar cannot admit a Lanczos tensor by virtue of the absence of a Lanczos potential tensor for the Riemann tensor \cite{RieNoLanc}.
Then how is the above-mentioned encoding of the gravitational information  into the Lanczos tensor  realized? Perhaps some another unknown feature of the tensor is at work, further studies may unearth it.

It may however be noted that the force analogy of the Lanczos tensor has limitations. The analogy does not imply that a geometric theory of gravitation, for instance GR, would mimic the Newtonian action-at-a-distance concept (which requires an infinite speed of propagation of gravity) for explaining observations. The domain of applicability of the force analogy of the Lanczos tensor is the same as that of the potential analogy of the metric tensor, i.e. the slow motion and the weak field  limit of gravity.

\section{Gravitational Wave and its Energy-Momentum}

Existence of gravitation waves is one of the most important features of GR, which was first predicted by Einstein.  In the light of the direct detection of the gravitational waves from binary mergers, it is now possible to extract information about matter and spacetime under extreme conditions.  

On the theoretical side however, there is no general agreement on
the description of the energy-momentum carried by the gravitational waves. The reason is that there is no existence of a localizable (i.e., tensor-represented)  point-wise gravitational energy-momentum density in GR.
The use of the non-covariant energy-momentum pseudo-tensors totally obscures the analysis of the subject.  Besides their non-uniqueness,  they can be annihilated at will at any given spacetime point by the right choice of coordinates. Thus, in the framework of pseudo tensor-formulation of the energy-momentum, one concludes that the energy carried by a gravitational wave is coordinate dependent.
Thus, the pseudo tensor-formulation of the energy-momentum density of the gravitational waves is though practically useful, it is not satisfactory from the theoretical point of view. We need a fully covariant tensorial formulation of this quantity with the requirement that it reduces to the pseudo tensor-formulation in a weak field.

In our search for a covariant tensorial formulation of the energy-momentum of the gravitational waves, we get a clue from the results obtained in the preceding section that the Lanczos tensor is a relativistic analogue of the Newtonian force. Hence its square has the same dimensions as that of the energy density (in geometric units). This means that, following the gravitation-electrodynamics correspondence,  an energy-momentum tensor can be formulated in terms of the Lanczos tensor, along the lines of the energy-momentum tensor of the electromagnetic field, in which the Lanczos tensor appears quadratically.
It is shown in the following that the so framed energy-momentum tensor gives the same results in the linearized theory as do the conventional pseudo tensor-based formulations for the energy-momentum of the gravitational waves.

\subsection{Linearized Gravity and Isaacson's Energy-Momentum Pseudo Tensor}

A simple wave equation in GR, is a linearized approximation of Einstein's equation where the velocities are small and the gravitational fields are weak, so that the spacetime metric is given by (\ref{eq:weak}).
Einstein's equation,  in a Ricci-flat spacetime, then yields 
\be
\partial^\kappa \partial_\kappa \bar{h}_{\mu\nu}=0 ~~~~~ {\rm with} ~~~~~ \bar{h}_{\mu\nu} \equiv h_{\mu\nu}-\eta_{\mu\nu}h/2, \label{eq:waveE}
\ee
which is a spin-2 field equation. 
The linearized Einstein equation takes the simple form (\ref{eq:waveE}) only if the Lorenz gauge conditions $\partial_\mu \bar{h}^{\mu\nu}=0$ are satisfied. This requirement makes the metric perturbation in equation (\ref{eq:waveE}) look like a transverse wave. By imposing the Lorenz gauge conditions, we have reduced the 10 independent components of the symmetric tensor $\bar{h}_{\mu\nu}$ to six. Since there are really only two independent components of the Riemann tensor in the present case, the remaining freedom is used to choose a gauge in which the perturbation becomes traceless ($\bar{h}\equiv \bar{h}_{\mu\nu}\eta^{\mu\nu}=0=h\equiv  h_{\mu\nu}\eta^{\mu\nu}$). This defines the transverse and traceless (TT) gauge leaving the only independent components of $\bar{h}_{\alpha\beta}$ ($=h_{\alpha\beta}$) as
\begin{subequations}
\label{eq:sSolutions}
\begin{align}
h_{11} & =-h_{22} \equiv h_+,       \label{eq:h11}\\
h_{12} & =h_{21} \equiv h_\times.    \label{eq:h12}
\end{align}
\end{subequations}
In this case, the plane wave solution of  equation (\ref{eq:waveE}) traveling along the $z=x^3$-direction can be written as
\be\label{eq:WaveSol} 
h_{\mu\nu}=A_{\mu\nu}\cos\big(\omega(t-z)\big),   
\ee
where $A_{\mu\nu}$ is the constant amplitude of the wave and $\omega$ its angular frequency.

As has been mentioned earlier, the description of the energy-momentum carried by the gravitational waves is still a disputed topic.  
Generally, the energy-momentum tensor for gravitational waves  is obtained from Isaacson's stress-energy pseudo-tensor
$\tau_{\mu\nu}$ \cite{EnergyGW} by averaging the squared gradient of the wave field over a length much larger than
the typical gravitational wavelength:
\be
\tau_{\mu\nu}=\frac{1}{32\pi}\langle \nabla_\mu h_{\alpha\beta} \nabla_\nu h^{\alpha\beta}\rangle.  \label{eq:tauEM}
\ee
In the above-defined TT gauge, the non-vanishing components of $\tau_{\mu\nu}$ are
\be
\tau_{00}=-\tau_{03}=-\tau_{30}=\tau_{33}=\frac{1}{16\pi}\langle(\dot{h_+})^2+(\dot{h_\times})^2\rangle,\label{eq:tauEMvalues}
\ee
where an overhead dot denotes derivative with respect to $t$.

\subsection{Lanczos Wave and its Energy-Momentum}

As has been mentioned earlier, there exists a perfect correspondence between electrodynamics and gravitation and various analogues of electromagnetic phenomena have been discovered in gravitation \cite{analogies}.
Electrodynamics admits a wave equation for the electromagnetic 4-potential. Note that this is so not only in flat spacetime, but also in the curved one.
It is already known (see, for example, \cite{Wald, IJGMMP}) that  a Killing vector field $A^\mu$ in a Ricci-flat  spacetime plays the role of the electromagnetic 4-potential and the source-free Maxwell equations, in Lorenz gauge ($A^\kappa_{~~;\kappa}=0$), reduce to the homogeneous wave equation
\be
\nabla^\kappa \nabla_\kappa A_\mu  =0.\label{eq:Max}
\ee
As  the Lanczos potential constitute a gravitational analogue of electromagnetic 4-potential, we expect a similar equation for $ L_{\mu\nu\sigma}$ if the correspondence between electrodynamics and gravity has any real meaning. Remarkably, the Lanczos potential  indeed satisfies a homogeneous wave equation
\be
\nabla^\kappa\nabla_\kappa L_{\mu\nu\sigma} =0, \label{eq:waveL}
\ee
in Lanczos gauge, in any Ricci-flat spacetime \cite{review, Edgar} ({\it not} necessarily in a weak field), strengthening our belief in the correspondence.
Akin to the wave equation (\ref{eq:waveE}), this simple and beautiful exact analytical solution is admitted in a Ricci-flat spacetime in any metric theory of gravity formulated in a 4-dimensional pseudo-Riemannian manifold.

This cannot be just a formal accident and must have some deeper meaning.
We know that a plane wave describes, from the quantum point of view, a collection of massless quanta - photons - with  helicity $\pm1$ for electrodynamics.
Hence, the electrodynamics-gravity correspondence, in view of equations (\ref{eq:Max}) and (\ref{eq:waveL}), insinuates that the plane wave (\ref{eq:waveL}) transports the excitations
of a massless field. That is, it describes at  the quantum level, a collection of massless quanta -
gravitons - with helicity $\pm2$ mediating gravity. Hence, its energy-momentum must necessarily be represented by a trace-free symmetric tensor, which is expected to reduce to the pseudo tensor-formulation (\ref{eq:tauEM}) in a weak field.
This might be helpful to develop an effective description of the physics of graviton which
captures some quantum effects but is otherwise based on classical concepts. This would also fit nicely with the earlier findings that the Lanczos potential is impregnated with various signatures of quantum physics \cite{Lanczos, vishwa_CQG, N-R}.

Supplemented with the novel insight that the energy-momentum tensor for the wave (\ref{eq:waveL}) must be symmetric and trace-free, and the earlier-noted cue that this tensor should have $L_{\mu\nu\sigma}$ appearing quadratically,
we define the following tensor along the lines of the energy-momentum tensor of the electromagnetic field $(\overset{\rm EM}{T}_{\alpha\beta}= F^{~\sigma}_{\alpha} F_{\beta\sigma} + *F^{~\sigma}_{\alpha} ~{*F}_{\beta\sigma})$:
\be
T_{\alpha\beta\gamma\delta} = \kappa(L^{~~\sigma}_{\alpha~~~\gamma}~ L_{\beta\sigma\delta} + *L^{~~\sigma}_{\alpha~~~\gamma}~ {*L}_{\beta\sigma\delta}),\label{eq:BRL4}
\ee
which is though not symmetric in all pair of indices. [A dimensional constant $\kappa$ has been inserted in equation (\ref{eq:BRL4}), which is given in terms of $G$ and $c$ in ordinary units.] A symmetric tensor can be obtained from (\ref{eq:BRL4}) by contracting over the last pair of indices of $T_{\alpha\beta\gamma\delta}$, giving
\be
T_{\alpha\beta} = \kappa \left(L^{~~\sigma\rho}_{\alpha}~ L_{\beta\sigma\rho} + *L^{~~\sigma\rho}_{\alpha}~ {*L}_{\beta\sigma\rho}\right),\label{eq:BRL}
\ee
which is indeed symmetric and trace-free (even if $L_{\mu\nu\sigma}$ does not satisfy the trace-free condition) like the energy-momentum tensor of the electromagnetic field. 
By evaluating the duals, equation (\ref{eq:BRL}) can be recast in the form
\be
T_{\alpha\beta} =\kappa \left(2 L^{~~\sigma\rho}_{\alpha}~ L_{\beta\sigma\rho} - \frac{1}{2}g_{\alpha\beta} L^{\kappa\sigma\rho}L_{\kappa\sigma\rho}\right).\label{eq:EM-Lan}
\ee
In order to compare this tensor-based definition of the energy-momentum of the gravitational waves with the traditional pseudo tensor one, let us calculate it in the weak gravity case (\ref{eq:weak}). But before that, let us first check if the wave equation (\ref{eq:waveL}) is admitted in a weak field. That is, if the Lanczos gauge conditions are admitted in a weak gravity field.
For this purpose, we calculate the trace and the divergence of the Lanczos tensor field given by (\ref{eq:weakL}) in a weak field. They lead to
\begin{subequations}
\begin{align}
& L^{~~\lambda}_{\mu~~\lambda} =\frac{1}{4} \bar{h}^{~\lambda}_{\mu~,\lambda},\\
& L^{~~~\lambda}_{\mu\nu~~,\lambda} =\frac{1}{4}\left(\bar{h}^{~\lambda}_{\mu~,\lambda,\nu} - \bar{h}^{~\lambda}_{\nu~,\lambda,\mu} \right).
\end{align}
\end{subequations}
Thus the tensor $L_{\mu\nu\sigma}$ does admit the Lanczos gauge conditions and hence the wave equation (\ref{eq:waveL}) in the TT gauge in a weak field. 
We can now come back to equation (\ref{eq:EM-Lan}). In order to calculate it in a weak field, we use the definition  of $L_{\mu\nu\sigma}$ given by (\ref{eq:weakL}). In  the TT gauge, this evaluates the following two terms appearing in (\ref{eq:EM-Lan}) as
\begin{subequations}
\label{eq:T12}
\begin{align}\nonumber
L^{~\sigma\rho}_\alpha L_{\beta\sigma\rho} 
& =\frac{1}{16}\big( 2 h_{1 1, \alpha }~  h_{1 1, \beta} + 2  h_{1 2, \alpha}~  h_{1 2, \beta}+ h_{\alpha 1, 3}~ h_{\beta 1, 3}+h_{\alpha 2,3}~ h_{\beta 2, 3}\\ 
&~~~~~~~~~~~ - h_{\alpha 1,0}~ h_{\beta 1,0}- h_{\alpha 2,0}~h_{\beta 2,0}  \big), \label{eq:T1} \\
L^{\kappa\sigma\rho}L_{\kappa\sigma\rho}
&=\frac{1}{4}\big( (h_{1 1,0})^2+(h_{1 2,0})^2 - (h_{1 1,3})^2-(h_{1 2,3})^2 \big). \label{eq:T2}
\end{align}
\end{subequations}
By virtue of $\partial h_{\mu\nu}/\partial z=-\partial h_{\mu\nu}/\partial t$, which readily follows  from solutions (\ref{eq:WaveSol}), $L^{\kappa\sigma\rho}L_{\kappa\sigma\rho}$ vanishes identically and hence the second term on the r.h.s. of (\ref{eq:EM-Lan}) does not contribute anything to $T_{\alpha\beta}$. Similarly, by the use of this identity, equation 
(\ref{eq:T1}) reduces to
\[
L^{~\sigma\rho}_\alpha L_{\beta\sigma\rho} =\frac{1}{8}\big(\partial_\alpha h_+  ~\partial_\beta h_+ +  \partial_\alpha h_\times ~\partial_\beta h_\times \big),
\]
giving
\be
T_{\alpha\beta} =\frac{\kappa}{4}\big(\partial_\alpha h_+  ~\partial_\beta h_+ +  \partial_\alpha h_\times ~\partial_\beta h_\times \big).
\ee
Thus the only non-vanishing components of $T_{\alpha\beta}$ are obtained as
\be
T_{00}=-T_{03}=-T_{30}=T_{33}=\frac{\kappa}{4}\left((\dot{h}_+)^2+(\dot{h}_\times)^2  \right). \label{eq:LEMvalues}
\ee
Remarkably, these values do match with their pseudo-tensor counterparts given by (\ref{eq:tauEMvalues}) for $\kappa=1/4\pi$. But, here they emerge from a fully covariant tensorial definition, which is intuitively appealing.

One can guess, how this energy-momentum tensor interacts with that of the dynamical matter fields when the gravitational wave travels through matter.  There will necessarily be an exchange of energy and momentum between matter and the Lanczos field, and what should be conserved is not the energy and momentum of the matter alone, but that of the sum of all the fields contributing to the structure of spacetime.

\section{Matching of the Lanczos Tensor at the Surface of a Gravitating Body}

In the general-relativistic stellar modelling, the exterior and the interior spacetime geometries are glued together at the hypersurface by using the Israel-Darmois junction conditions \cite{Poisson}. This requires that the extrinsic curvature and the induced metric on the hypersurface must be the same on both sides of the hypersurface. Well-known examples are: 
\begin{enumerate}
\item  
The Schwarzschild interior metric (\ref{eq:sch-Int}), representing the spacetime inside an idealized star of constant density, is joined with the Schwarzschild exterior metric (\ref{eq:sch}). 
\item
 The dynamical cosmological interior (\ref{eq:RW}), modeling an idealized collapsing star of spatially constant density, is joined with the Schwarzschild static  exterior metric (\ref{eq:sch}).
\end{enumerate}
Why are the junction conditions limited to matching only the extrinsic curvature and the induced metric at the boundary? Because the union of the two metrics form a valid solution to Einstein field equation in this case.
However in a geometric theory of gravitation wherein gravity appears through the curvature, we should be able to calculate the gravitational effect on a test mass placed at the boundary, from either of the two metrics. Should we then expect the curvature of the spacetimes too to match on the both sides of the boundary? This expectation is however not fulfilled in the above examples. 
In both these examples, the interiors are given by the conformally flat spacetimes giving vanishing Weyl tensor. Whereas the exteriors are Ricci-flat spacetimes. Thus, the Riemann tensor cannot be matched at the boundary, as is ascertained by equation (\ref{eq:R-W-R}). 

How else can we measure the gravitational effect experienced by the test mass? In terms of a quantity which is an analogue of Newtonian gravitational force - the Lanczos tensor! Interestingly, the Lanczos tensor indeed matches at the boundary, as we shall see in the following.

\subsection{Matching the Lanczos Tensor on an Idealized Static Star}
\label{sec:StaticStar}

As has been mentioned earlier, the line element given by (\ref{eq:sch-Int}), i.e.
\[
ds^2=\frac{1}{4}\left(3\sqrt{1-\frac{r_0^2}{R^2}}-\sqrt{1-\frac{r^2}{R^2}}\right)^2 dt^2-\frac{dr^2}{1-r^2/R^2}-r^2 d\theta^2-r^2\sin^2\theta ~d\phi^2,
\]
represents the static and isotropic spacetime inside a non-rotating star of radius $r_0$, composed of an incompressible perfect fluid of constant density $\rho_0$ and variable pressure $p$. This spacetime is glued with the Schwarzschild static exterior spacetime given by the line element (\ref{eq:sch}), i.e.
\[
ds^2=\left(1-\frac{2m}{r}\right) dt^2-\frac{dr^2}{(1-2m/r)}-r^2d\theta^2-r^2\sin^2\theta ~d\phi^2.
\]
It is well-known that these interior and exterior spacetimes have a perfect matching at the boundary $r=r_0$ (see for example, \cite{JVN}).

We have already calculated the Lanczos tensor for these interior and exterior spacetimes in sections \ref{sec:Lsch-Int} and \ref{sec:LSch} respectively.
Let us first consider the values of the tensor in these spacetimes satisfying the gauge condition of divergence-freeness (\ref{eq:conditionb}). These are given by equations (\ref{eq:Lan1-sch-Int}) and (\ref{eq:Lan1-sch}) respectively for the interior and the exterior spacetimes:
\[
 \overset{\rm int}{L}_{100}=\frac{5 r\left[ 1-\frac{r^2+9r_0^2}{10R^2} -\frac{3}{5}\sqrt{\left(1-\frac{r^2}{R^2}\right)\left(1-\frac{r_0^2}{R^2}\right)}\right]}{2 R^2\left[ \left(1-\frac{r^2}{R^2}\right)-3\sqrt{\left(1-\frac{r^2}{R^2}\right)\left(1-\frac{r_0^2}{R^2}\right)}\right]}
~~~~~~~~\leftrightarrow ~~~~~~~~
\overset{\rm ext}{L}_{100}=-\frac{m}{r^2},
\]
which become equal at the boundary $r=r_0$ by virtue of $R^2=3/8\pi \rho_0$ and realizing that the total mass of the interior $=4\pi r_0^3 \rho_0/3$ constitutes the source mass $m$ for the exterior spacetime.
The same is true for the Lanczos tensor satisfying both Lanczos gauge conditions (\ref{eq:conditiona}) and (\ref{eq:conditionb}). In this case, the respective values of the tensor are given by equations  (\ref{eq:Lan2-sch-Int}) and  (\ref{eq:Lan2-sch}) respectively for the interior and  the exterior spacetimes. At the boundary  $r=r_0$, these values reduce to
\[
\left.\begin{aligned}
&  \overset{\rm int}{L}_{100}= -\frac{ r_0}{3 R^2}
~~~~~~~~~~~~~~~~~~~~~~~~~~~~\leftrightarrow ~~~~~~~~~
\overset{\rm ext}{L}_{100}=-\frac{2m}{3r_0^2},\\
& \overset{\rm int}{L}_{122}= -\frac{ r_0^3}{6 R^2\left(1-\frac{r_0^2}{R^2}\right)}
~~~~~~~~~~~\leftrightarrow ~~~~~~~~~~
\overset{\rm ext}{L}_{122}=-\frac{m}{3(1-2m/r_0)},\\
& \overset{\rm int}{L}_{133}=-\frac{ r_0^3 \sin^2\theta}{6 R^2\left(1-\frac{r_0^2}{R^2}\right)}
~~~~~~~~~~~\leftrightarrow ~~~~~~~~~~
\overset{\rm ext}{L}_{133}=-\frac{m\sin^2\theta}{3(1-2m/r_0)},
 \end{aligned}
 \right.
\]
which coincide by virtue of  $1/R^2=2m/r_0^3$ as before. Thus the Lanczos tensor does  match smoothly at the common boundary despite the Weyl tensor differing in the two regions. This strengthens our observation that the Lanczos tensor is a relativistic analogue of the Newtonian gravitational force.

It however appears that the interior (\ref{eq:sch-Int}) considered in these examples, which was derived by Schwarzschild to have a mathematically simple solution,  is not physically viable.
As this solution assumes  a static sphere of matter consisting of a perfect fluid of constant density $\rho$ but a variable pressure $p(r)$ that vanishes at the boundary, the speed of sound $=\sqrt{dp/d\rho}$ becomes infinite  in the fluid. 
In order to avoid this situation, let us consider, in the following section, an interior composed of a dust cloud with vanishing pressure, which gives rise to a collapsing interior.

\subsection{Matching the Lanczos Tensor on a Collapsing Star}
\label{sec:CollapsingStar}

The solution of Einstein field equation for a homogeneous dust cloud of spherical symmetry, leads to the well-known problem of  spherical gravitational collapse. The spacetime inside the collapsing sphere is represented by the dynamical R-W metric (\ref{eq:RW}) with the spatially closed case, which  is matched with the Schwarzschild static exterior (\ref{eq:sch}). This  problem of  the imploding  dust ball was considered by Datt  in 1938 \cite{collapse1} and by Oppenheimer \& Snyder in 1939 \cite{collapse2}. For the sake of completeness, we have summarized the formal aspects of the theory in the Appendix.

Let us note that unlike the case studied in the preceding section, here the matching of the two spacetime regions take place  at a moving boundary between the dynamical cosmological interior (\ref{eq:RW}) and the static Schwarzschild exterior (\ref{eq:schn}).
In order to study the matching of the accompanied Lanczos potentials in the two regions, we first consider the potentials with single non-vanishing components.  These are given by equations (\ref{eq:Lan2-RW}) and (\ref{eq:Lan1-sch}) for the interior and the exterior spacetimes respectively:
\be
 \overset{\rm int}{L}_{100}=\frac{\kappa r S^n}{\sqrt{1-\alpha r^2}}~~~~~~~~~~~~~\leftrightarrow ~~~~~~~~~~~~~
\overset{\rm ext}{L}_{100}=-\frac{m}{{\bar{r}}^2}. \label{eq:potentials0}
\ee
As the coordinates used in the two spacetime regions are different, the accompanied potentials shown in (\ref{eq:potentials0}) also have different coordinates. In order to make the comparison possible, we use the transformation (\ref{eq:trans}) to bring the potentials in a single (comoving) coordinate system. This gives
\be
 \overset{\rm int}{L}_{100}=\frac{\kappa r S^n}{\sqrt{1-\alpha r^2}}~~~~~~~~~~~~~\leftrightarrow ~~~~~~~~~~~~~
\overset{\rm ext}{L}_{100}=-\frac{m}{r^2 S^2}.\label{eq:potentials1}
\ee
Let us recall that the coordinates $r,\theta, \phi$ are the comoving coordinates of the spacetime (\ref{eq:RW}) and hence are independent of time. Hence, a continuous contraction in the surface (boundary) of the ball can be brought about only through the function $S=S(t)$, which is a decreasing function of the proper time $t$, as is indicated by (\ref{eq:dyn}). This implies that both the potentials in (\ref{eq:potentials1}) must have the same $S$-dependence to ensure a proper matching and  continuity across the boundary throughout the collapse. This uniquely determines the parameter $n$ appearing in (\ref{eq:Lan2-RW}) as $n=-2$. Interestingly, this value of $n$ brings the potential (\ref{eq:Lan2-RW}) in the same (divergence-free)  gauge, as is (\ref{eq:Lan1-sch}), which is clear from equation (\ref{eq:divergence}). Thus the final form of the potentials to be matched is 
\be
 \overset{\rm int}{L}_{100}=\frac{\kappa r}{\sqrt{1-\alpha r^2}~S^2}~~~~~~~~~~~~~\leftrightarrow ~~~~~~~~~~~~~
\overset{\rm ext}{L}_{100}=-\frac{m}{r^2 S^2},\label{eq:potentials2}
\ee
which indeed match at the boundary $r=r_{\rm b}$ throughout the collapse as the matching becomes independent of $S(t)$, and hence independent of time. This determines the constant $\kappa$ uniquely as $\kappa=-\alpha\sqrt{1-\alpha r_{\rm b}^2}/2$ by virtue of $\alpha= 8\pi\rho_0/3=2m/r_{\rm b}^3$,
thus resulting in a smooth fit which appropriately accounts for continuity in the Lanczos potentials across the sphere boundary.

Let us now consider the matching of the Lanczos tensor having more than one non-vanishing components in the two regions of spacetime, for instance (\ref{eq:Lan4-RW}) versus (\ref{eq:Lan2-sch}): 
\be
\left.\begin{aligned}
& \overset{\rm int}{L}_{100}=\frac{\kappa_1 r}{\sqrt{1-\alpha r^2}S^2}
~~~~~~~~\leftrightarrow ~~~~~~~~~
\overset{\rm ext}{L}_{100}=-\frac{2m}{3\bar{r}^2}=-\frac{2m}{3r^2 S^2},\\
& \overset{\rm int}{L}_{122}=\frac{\kappa_2 r^3}{\sqrt{1-\alpha r^2}}
~~~~~~~~~~~~\leftrightarrow ~~~~~~~~~~
\overset{\rm ext}{L}_{122}=-\frac{m}{3(1-2m/\bar{r})}=-\frac{m}{3\left(1-\frac{2m}{rS}\right)},\\
& \overset{\rm int}{L}_{133}=\frac{\kappa_2 r^3\sin^2\theta}{\sqrt{1-\alpha r^2}}
~~~~~~~~~~~\leftrightarrow ~~~~~~~~~~
\overset{\rm ext}{L}_{133}=-\frac{m\sin^2\theta}{3(1-2m/\bar{r})}=-\frac{m\sin^2\theta}{3\left(1-\frac{2m}{rS}\right)}.
 \end{aligned}
 \right.  \label{eq:PotDyn}
\ee
An inspection reveals that only the first component (i.e., $L_{100}$) of the interior and exterior potentials diverge at $S=0$, but the rest two components do not.  
While the last two interior ones remain fixed on the boundary, the corresponding exterior ones vanish as $S\rightarrow0$. 
Hence they do not seem to match in this form. Let us recall that in the absence of sufficient conditions of plausibility that should be imposed to give a unique value of the  tensor $L_{\mu\nu\sigma}$, there remain redundant degrees of freedom in it and hence some of its values (in some arbitrary gauge) may not be physically meaningful. However, it would be interesting to see under what conditions the values given in (\ref{eq:PotDyn}) can match. 
In this context, one can check that if the collapsing $S$ can halt at a non-vanishing minimum of $S$, say $S_{\rm min}$, then all the three components of the tensor match at the boundary after the ball settles to the minimum radius.
This determines $\kappa_1=-\alpha\sqrt{1-\alpha r_{\rm b}^2}/3$ and $\kappa_2=-\frac{\alpha}{6}\frac{\sqrt{1-\alpha r_{\rm b}^2}}{(1-\alpha r_{\rm b}^2/S_{\rm min})}$.

The above result does not mean that the Lanczos tensor acts as a repulsive agent that can halt the collapse. Rather it provides an indication that there must exist an era when the quantum gravity-effects take over at some $S_{\rm min}$, arresting the collapse and avoiding the singularity.
There have been various claims that the singularities of GR can be avoided  by using quantum effects.
Although a self-consistent theory of quantum gravity remains elusive, there is a general agreement that removal of classical gravitational singularities is not only a crucial conceptual test of any theory of  quantum gravity but also a prerequisite for any reasonable theory of quantum gravity.
Let us note that the existence of an $S_{\rm min}$ is also not inconsistent with the matching discussed in equation (\ref{eq:potentials2}).

\section{Concluding Remarks}

Einstein's revolutionary insight - the local equivalence of gravitation and inertia - paved the way 
to the lofty goal of the geometrization of gravitation by introducing (pseudo) Riemannian geometry wherein 
the Riemann-Christoffel curvature tensor plays the central role for the unfolding of
gravity in a four-dimensional spacetime manifold.
However, it is not the full curvature tensor which takes part in the field equations of gravitation. Rather, only its trace - the Ricci tensor and Ricci scalar - appear in Einstein's equation. 
It is the Lanczos tensor which fulfills this gap by generating the trace-free part of the Riemann tensor -  exactly those components of the tensor which are missing in Einstein's equation. 

Despite its importance and novelty, the Lanczos tensor has not been paid proper attention as it deserves.  
One of the reasons of its obscurity  is that it is not clear what it represents physically.
By deriving expressions for the tensor in some particularly chosen spacetimes, we have
attempted to find its physical meaning and an adequate interpretation.
It has been shown that the especial property of the Weyl tensor - its conformal invariance - imposes additional constraints on the Lanczos tensor, not realized before.

The traditional viewpoint on the Lanczos tensor comes from the well-studied correspondence between electrodynamics and gravitation. Thence appears the perspective that the Lanczos tensor is a gravitational analogue of the electromagnetic 4-potential since it generates Weyl tensor differentially.
A complementary perspective on the tensor has been discovered which emerges from the weak field and low velocity limit of GR:  One may interpret the Lanczos tensor in a curved spacetime as a relativistic, tensorial analogue of the Newtonian gravitational force. Additionally, the Minkowskian Lanczos tensor appears as part and parcel of the Lanczos tensor field of any curved spacetime in a weak field and hence can be interpreted as the `weight' of the spacetime, providing a substantive status to it.
This emphasizes the importance of the Lanczos tensor in terms of a multi-featured quantity, which is also corroborated by its applications devised in the preceding sections.

The relation of this tensor to the energy-momentum of the gravitational waves, and its matching on the surface of a gravitating body give added proof of its fundamental significance.
Interestingly, the matching of the potential at the boundary of a collapsing mass provides an indication of the existence of  quantum gravity-effects avoiding the singularity. This finding attributes a quantum signature to the Lanczos potential, which appears consistent with some other studies revealing a quantum character of the potential tensor.

It appears that the theory of the Lanczos tensor field contains surprises and insights which have only just started to reveal.
The performed study has introduced a new domain of applicability of this tensor field. It reveals that the tensor field, which requires for its construction  merely the basic concepts of spacetime, is impregnated with interesting physical properties and opens up  a new research avenue with rich prospects.

 \renewcommand{\theequation}{A-\arabic{equation}}
  \setcounter{equation}{0}  
  \section*{Appendix: {\it Dutt-Oppenheimer-Snyder model of collapsing dust cloud $-$ A Brief Review}}

This is a simple model of  spherical gravitational collapse of a dust ball, which has considerable methodological importance. 
Let us consider a non-rotating ball of electrically neutral matter having a spherical symmetry in its physical parameters such as density and pressure. Let us ignore the effect of pressure as an opposing agency to gravity. 
This is a reasonable assumption for the stars in which the fusion process has nearly ceased and hence little radiation pressure remains.
Moreover, the pressure generated during collapse is not adequate to halt the collapse.

By solving Einstein's field equation for a dynamic, spherically symmetric line element with the energy-momentum tensor for a pressure-less perfect fluid, it has been shown that
 such a ball of pressure-less dust of uniform but time-dependent density $\rho$, must contract without limit under self gravitation \cite{H-N}. 
The spacetime interior to the dust ball turns out to be given by the spatially closed R-W line element (\ref{eq:RW}):
\[
ds^2=dt^2-S^2(t)\left[\frac{dr^2}{1-\alpha r^2}+r^2(d\theta^2+\sin^2\theta ~d\phi^2)\right],
\]
where $\alpha\equiv 8\pi\rho_0/3$ with $\rho_0$ being the initial density of the dust particles measured by the observer riding the surface of the collapsing ball. Clearly such an observer uses comoving 
 coordinates $r,\theta,\phi$ and thus measures proper time $t$. He thus measures the total mass of the ball as $m=4\pi \rho_0 r_{\rm b}^3/3$, with $r=r_{\rm b}$ being the radial coordinate of the boundary of the dust ball. The dynamics of the collapse is regulated by the dynamics of the function $S(t)$ through
\be
\dot{S}(t)=-\sqrt{\alpha}\left[\frac{1}{S(t)}-1\right]^{1/2},\label{eq:dyn}
\ee
obtained by normalizing the radial coordinate so that $S(0)=1$, and by assuming that the dust particles are at rest initially in the chosen coordinates, so that $\dot{S}(0)=0$. Equation (\ref{eq:dyn}) can easily be integrated in the following parametric form:
\be
S=\frac{1}{2}(1+\cos\psi), ~~~~~~~ t=\frac{\psi+\sin\psi}{2\sqrt{\alpha}}.
\ee
This indicates that $S=0$ when $\psi=\pi$. Thus the observer riding the surface of the ball, registers it collapsing into the centre  with $S=0$ in proper time 
$
\frac{\pi}{2\sqrt{\alpha}}.
$
However,  a distant external observer measures a different time interval for this process.  
Let us note that the spacetime exterior to the boundary is described by the Schwarzschild line element (\ref{eq:sch})
\be
ds^2=\left(1-\frac{2m}{\bar{r}}\right) d\bar{t}^2-\frac{d\bar{r}^2}{(1-2m/\bar{r})}-\bar{r}^2(d\bar{\theta}^2+\sin^2\bar{\theta} ~d\bar{\phi}^2),\label{eq:schn}
\ee
 as may be inferred from the Birkhoff theorem.
As this line element is not in the Gaussian normal form as in (\ref{eq:RW}), we have departed from the earlier notation used for coordinates in (\ref{eq:sch}). 
It can now be shown \cite{H-N} that a light signal emitted in the radial direction at time ${\bar{t}_1}$ from a point $\bar{r}={\bar{r}_1}$ on the surface of the collapsing ball, will arrive at an external observer located at $\bar{r}={\bar{r}_2}$ at time ${\bar{t}_2}$ given by
\be
{\bar{t}_2}={\bar{t}_1}+\int^{\bar{r}_2}_{\bar{r}_1} \left(1-\frac{2m}{x}\right)^{-1}dx,
\ee
indicating that ${\bar{t}_2}\rightarrow \infty$ as ${\bar{r}_1}\rightarrow 2m$ (Schwarzschild radius). Thus the collapse of the ball to the Schwarzschild radius, appears to an outside observer to take an infinite time. Hence, the collapse to $S=0$ is utterly unobservable from outside.
This classical, unavoidable collapse of all the particles of the ball into the centre, is a true singularity of spacetime, as $S=0$ signifies a state wherein the entire space shrinks to zero volume with the density going to infinity. 

As the coordinates used in the interior and the exterior line elements (\ref{eq:RW}) and  (\ref{eq:schn}) are different, we have to convert one into the other  in order to facilitate the matching of the two line elements at the boundary of the dust ball. 
Comparing the angular parts of the line elements (\ref{eq:RW}) and  (\ref{eq:schn}), it is evident that the comoving R-W coordinates in (\ref{eq:RW}) and those in (\ref{eq:schn}) are related by
\be
\bar{r}=r S(t), ~~~~ \bar{\theta}=\theta, ~~~~ \bar{\phi}=\phi.\label{eq:trans}
\ee
By the use of these transformations, the interior line element (\ref{eq:RW}) can be expressed in Schwarzschild coordinates
$\bar{t}$, $\bar{r}$, showing that the interior and the exterior line elements fit continuously at the boundary $\bar{r}=r_{\rm b} S(t)$ by virtue of $\alpha\equiv 8\pi\rho_0/3=2m/r_{\rm b}^3$ \cite{H-N}.

\vspace{1cm}
\noindent
{\bf Acknowledgements:}
The author would like to thank Sanjeev V. Dhurandhar, Jayant V. Narlikar and Jos\'e M. M. Senovilla for many insightful discussions.
This paper has benefited from suggestions made by an anonymous reviewer.


\begin{thebibliography}{99}


\bibitem{Shoes} A. G. Riess, S. Casertano, W. Yuan, L. M. Macri and D. Scolnic, ``Large Magellanic cloud Cepheid standards provide a 1\% foundation for the determination of the Hubble constant and stronger evidence for physics beyond $\Lambda$CDM", {\it Astrophys. J.}, {\bf 876}, 85 (2019)

\bibitem{Lanczos} C. Lanczos,   ``The splitting of the Riemann tensor", {\it Rev. Mod. Phys.}, {\bf 34}, 379 (1962).

\bibitem{vishwa_CQG} R. G. Vishwakarma, ``Relativistic potential of Weyl: a gateway to quantum physics in the
presence of gravitation?", {\it Class. Quant. Grav.}, {\bf 37}, 065020 (2020).

\bibitem{Takeno} H. Takeno, ``On the spintensor of Lanczos", {\it Tensor, N.S.}, {\bf 14}, 103 (1964).

\bibitem{N-R} M. Novello and L. M. C. S. Ridrigues, ``A unified model for gravity and electroweak interactions", {\it Lett. Nuovo Cimento}, {\bf 43}, 292 (1985).

\bibitem{Edgar2000} S. B. Edgar and A. Hoglund, ``The Lanczos potential for Weyl-candidate tensors exists only in four dimensions",  {\it Gen. Relativ. Gravit.}, {\bf 32}, 2307 (2000).

\bibitem{Edgar-Senovilla}  S. B. Edgar and J. M. M. Senovilla, ``A local potential for the Weyl tensor in all dimensions", {\it Class. Quant. Grav.}, {\bf 21}, L133 (2004).

\bibitem{B-C} F. Bampi and G. Caviglia, ``Third-order tensor potentials for the Riemann and Weyl
tensors", {\it Gen. Relativ. Gravit.}, {\bf 15}, 375 (1983).

\bibitem{N-V} M. Novello and A. L. Velloso, ``The connection between the general observers and
Lanczos potential", {\it Gen. Relativ. Gravit.}, {\bf 19}, 1251 (1987).

\bibitem{Bonilla} J. L. Lopez-Bonilla, J. Morales and G. Ovando, ``A potential for the Lanczos spintensor in Kerr
geometry" {\it Gen. Relativ. Gravit.}, {\bf 31}, 413 (1999).

\bibitem{review}  S. B. Edgar and A. Hoglund, ``The Lanczos potential for the Weyl curvature
tensor: existence, wave equation and algorithms", {\it Proc. Roy. Soc. A}, {\bf 453},  835 (1997).

\bibitem{Wald} R. M. Wald,  {\it General Relativity}, (The University of Chicago Press, Chicago and London, 1984).

\bibitem{RieNoLanc} S. B. Edgar, `` Nonexistence of the Lanczos potential for the Riemann tensor in higher dimensions", {\it Gen. Relativ. Gravit.}, {\bf 26} 329 (1994); E. Massa and E. Pagani, ``Is the Riemann tensor derivable from a tensor
potential?", {\it Gen. Relativ. Gravit.}, {\bf 16} 805 (1994).

\bibitem{EnergyGW} R. Isaacson, ``Gravitational radiation in the limit of high frequency. II. Nonlinear terms and the effective stress tensor",   {\it Phys. Rev.}, {\bf 166}, 1272, (1968);
L. D. Landau and E. M. Lifshitz, {\it The classical theory of fields, course of theoretical physics}, vol 2, 4th edn.
(Butterworth-Heinemann, Oxford, 1975). 

\bibitem{analogies} R. Maartens and B. A. Bassett, ``Gravito-electromagnetism", {\it Class. Quant. Grav.}, {\bf 15}, 705 (1998);
L. Filipe Costa and C. A. R. Herdeiro, ``Gravitoelectromagnetic analogy based on tidal tensors", {\it Phys. Rev. D}, {\bf 78}, 024021 (2008);
R. G. Vishwakarma,  ``A new Weyl-like tensor of geometric origin", {\it J. Math. Phys.}, {\bf 59}, 042505  (2018).

\bibitem{IJGMMP}  R. G. Vishwakarma, ``A scale-invariant, Machian theory of gravitation and electrodynamics unified", {\it Int. J. Geom. Methods Mod. Phys.}, {\bf 15}, 1850178 (2018).

\bibitem{Edgar}
P. Dolan and C. W. Kim, ``The wave equation for the Lanczos potential: I",  {\it Proc. Royl. Soc. A}, {\bf 447}, 557 (1994).

\bibitem{Poisson} E. Poisson, {\it A  Relativist's Toolkit: the mathematics of black-hole mechanics}, (Cambridge University Press, 2004).

\bibitem{JVN} J. V. Narlikar, {\it An Introduction to Relativity}, (Cambridge University Press, 2010).

\bibitem{collapse1} B. Datt, ``Uber eine Klasse von Losungen der Gravitationsgleichungen
der Relativitat", {\it Z. Phys.}, {\bf 108}, 314 (1938).

\bibitem{collapse2} J. R. Oppenheimer and H. Snyder, ``On continued gravitational contraction", {\it Phys. Rev.}, {\bf 56}, 455 (1939).

\bibitem{H-N} F. Hoyle and J. V. Narlikar, ``On the avoidance of singularities in C-field cosmology", {\it Proc. Roy. Soc. A}, {\bf  278}, 465 (1964).

\end{thebibliography}
\end{document}